\documentclass[10pt,conference,compsocconf]{IEEEtran}
\usepackage{amsmath,amssymb,amsfonts}

\usepackage{graphicx}
\usepackage{mdframed}

\usepackage{float}

\usepackage{url}

\usepackage[utf8]{inputenc}

\makeatletter
\newcommand*\titleheader[1]{\gdef\@titleheader{#1}}
\AtBeginDocument{%
  \let\st@red@title\@title
  \def\@title{%
    \bgroup\normalfont\large\centering\@titleheader\par\egroup
    \vskip1.5em\st@red@title}
}
\makeatother

\usepackage{listings, xcolor}

\definecolor{verylightgray}{rgb}{.97,.97,.97}
\definecolor{verydarkgray}{rgb}{.12,.12,.12}

\definecolor{tealish}{HTML}{4EC9B0}
\definecolor{lightblue}{RGB}{51,153,255}

\lstdefinelanguage{Solidity}{
	keywords=[1]{anonymous, assembly, assert, balance, break, call, callcode, case, catch, class, constant, continue, constructor, contract, debugger, default, delegatecall, delete, do, else, emit, event, experimental, export, external, false, finally, for, function, gas, if, implements, import, in, indexed, instanceof, interface, internal, is, length, library, log0, log1, log2, log3, log4, memory, modifier, new, payable, pragma, private, protected, public, pure, push, require, return, returns, revert, selfdestruct, send, solidity, storage, struct, suicide, super, switch, then, this, throw, transfer, true, try, typeof, using, value, view, while, with, addmod, ecrecover, keccak256, mulmod, ripemd160, sha256, sha3}, 
	keywordstyle=[1]\color{lightblue}\bfseries,
	keywords=[2]{address, bool, byte, bytes, bytes1, bytes2, bytes3, bytes4, bytes5, bytes6, bytes7, bytes8, bytes9, bytes10, bytes11, bytes12, bytes13, bytes14, bytes15, bytes16, bytes17, bytes18, bytes19, bytes20, bytes21, bytes22, bytes23, bytes24, bytes25, bytes26, bytes27, bytes28, bytes29, bytes30, bytes31, bytes32, enum, int, int8, int16, int24, int32, int40, int48, int56, int64, int72, int80, int88, int96, int104, int112, int120, int128, int136, int144, int152, int160, int168, int176, int184, int192, int200, int208, int216, int224, int232, int240, int248, int256, mapping, string, uint, uint8, uint16, uint24, uint32, uint40, uint48, uint56, uint64, uint72, uint80, uint88, uint96, uint104, uint112, uint120, uint128, uint136, uint144, uint152, uint160, uint168, uint176, uint184, uint192, uint200, uint208, uint216, uint224, uint232, uint240, uint248, uint256, var, void, ether, finney, szabo, wei, days, hours, minutes, seconds, weeks, years},	
	keywordstyle=[2]\color{tealish}\bfseries,
	keywords=[3]{block, blockhash, coinbase, difficulty, gaslimit, number, timestamp, msg, gas, sender, sig, value, now, tx, gasprice, origin},	
	keywordstyle=[3]\color{violet}\bfseries,
	identifierstyle=\color{white},
	sensitive=false,
	comment=[l]{//},
	morecomment=[s]{/*}{*/},
	commentstyle=\color{gray}\ttfamily,
	stringstyle=\color{red}\ttfamily,
	alsoletter={)(\),;.\{\}[]},
	morestring=[b]',
	morestring=[b]"
}

\usepackage{xcolor}
\usepackage{algorithm}
\usepackage[noend]{algpseudocode}

\renewcommand{\Comment}[1]{\hfill //\emph{#1}}

\usepackage{etoolbox}
\newtoggle{IEEE_submission}

\newtoggle{detailed_version}
\iftoggle{IEEE_submission}{%
    \togglefalse{detailed_version}
}{%
}

\iftoggle{IEEE_submission}{%
}{%
    \usepackage{hyperref}
}

\iftoggle{IEEE_submission}{%
}{%
}

\title{Decentralized \& Collaborative AI on Blockchain}
\titleheader{2019 IEEE International Conference on Blockchain (Blockchain)}

\author{\IEEEauthorblockN{Justin D. Harris}
\IEEEauthorblockA{\textit{Microsoft Research} \\
Montreal, Canada \\
justin.harris@microsoft.com}
\and
\IEEEauthorblockN{Bo Waggoner}
\IEEEauthorblockA{\textit{Microsoft Research} \\
New York, USA \\
bwag@colorado.edu}
}

\begin{document}
\IEEEoverridecommandlockouts
\IEEEpubid{\makebox[\columnwidth]{978-1-7281-4693-5/19/\$31.00~\copyright2019 IEEE~DOI 10.1109/Blockchain.2019.00057 \hfill} \hspace{\columnsep}\makebox[\columnwidth]{ }}
\maketitle
\IEEEpubidadjcol

\begin{abstract}
Machine learning has recently enabled large advances in artificial intelligence, but these tend to be highly centralized.
The large datasets required are generally proprietary; predictions are often sold on a per-query basis; and published models can quickly become out of date without effort to acquire more data and re-train them.
We propose a framework for participants to collaboratively build a dataset and use smart contracts to host a continuously updated model.
This model will be shared publicly on a blockchain where it can be free to use for inference.
Ideal learning problems include scenarios where a model is used many times for similar input such as personal assistants, playing games, recommender systems, etc.
In order to maintain the model's accuracy with respect to some test set we propose both financial and non-financial (gamified) incentive structures for providing good data.
A free and open source implementation for the Ethereum blockchain is provided at \url{https://github.com/microsoft/0xDeCA10B}.
\end{abstract}

\begin{IEEEkeywords}
Decentralized AI, Blockchain, Ethereum, Crowdsourcing, Prediction Markets, Incremental Learning
\end{IEEEkeywords}

\section{Introduction}
We propose a framework for sharing and improving a machine learning model.
In this framework, anyone can freely access the model's predictions or provide data to help improve the model.
An important challenge is that the system must be robust and incentivize participation, but discourage manipulation.
Our framework is modular, and we propose and justify three example choices of ``incentive mechanisms'' with different advantages.

There exist several proposals to combine machine learning and blockchain frameworks.
In systems such as DInEMMo~\cite{Marathe2018DInEMMoDI}, access to the trained model is limited to a marketplace.
This allows contributors to profit based on a model's usage, but it limits access to those who can pay.
DanKu proposes storing already trained models in smart contracts for competitions, which does not allow for continual updating and collaborative training \cite{danku_protocol}.
In contrast, the goal of this work is to address the current centralization of artificial intelligence by sharing models freely.
Such centralization includes machine learning expertise, siloed proprietary data, and access to machine learning model predictions (e.g. charged on a per-query basis).

\subsection{Overview}
\begin{figure}[b!]
    \includegraphics[width=0.48\textwidth,trim=20 20 20 18,clip]
        {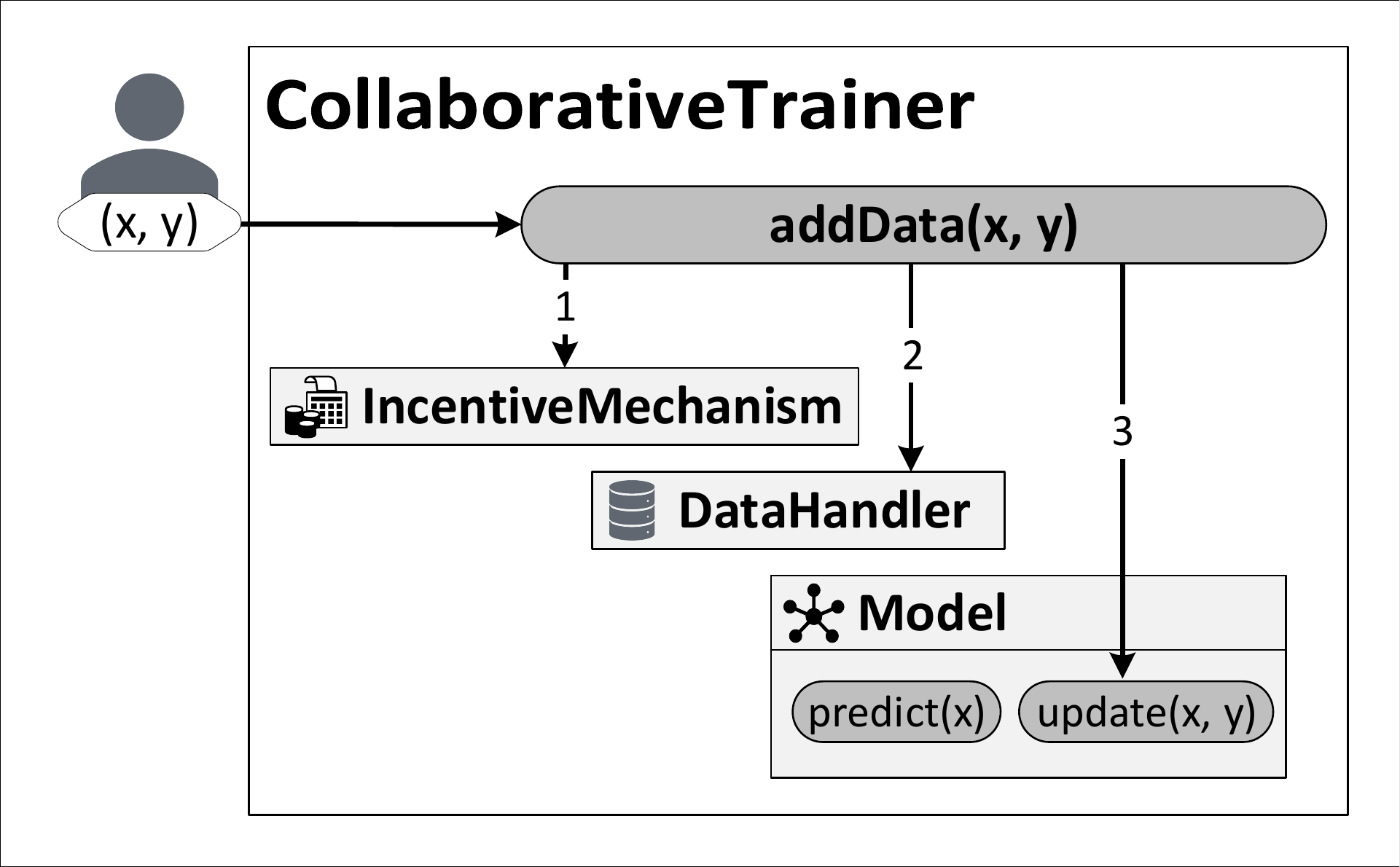}
    \caption{Adding data consists of 3 steps. (1) The IncentiveMechanism validates the transaction, for instance, in some cases a ``stake'' or monetary deposit is required. (2) The DataHandler stores data and meta-data onto the blockchain. This ensures that it is accessible for all future uses, not limited to this smart contract. (3) The machine learning model is updated according to predefined training algorithms. In addition to adding data, anyone can query the model for predictions, and the incentive mechanism may be triggered to provide users with monetary payments or virtual ``karma'' points.}
    \label{fig:architecture_flow}
\end{figure}

By leveraging advances in AI, prediction markets, and blockchain platforms, we can demonstrate the capabilities of a new framework to collect vast amounts of data, allow contributors to potentially profit, and host a shared machine learning model as a public resource.
The model can be collaboratively trained by many contributors yet remain open and free for others to use the model for inference.
This is accomplished with several configurable components:
\begin{itemize}
    \item the \emph{incentive mechanism}
    \item the \emph{data handler}
    \item the \emph{machine learning model}
\end{itemize}
A smart contract is created and initialized with choices for these components.
It then accepts ``add data'' actions from participants, with the incentive mechanism possibly triggering payments or allowing other actions.
Adding data involves validation from the incentive mechanism, storing in the data handler, and finally calling the \emph{update} method on the model's contract, as shown in Fig. \ref{fig:architecture_flow}.
Prediction is done \emph{off-chain} by calling the \emph{predict} function provided for convenience in the model's smart contract code.

The goal of our system is not for the creators to profit: the goal is to create valuable shared resources.
It is possible for the data contributors to profit financially (depending on the incentive mechanism) but this is mainly a result of mechanisms designed to penalize the contributors who submit bad data.

The dataset is also public because it can be found in the blockchain's transaction history or through emitted events (if this feature is available to the blockchain framework).
Collecting large datasets can be costly using typical crowdsourcing platforms such as Figure Eight (formerly known as Dolores Lab, CrowdFlower) and Amazon Mechanical Turk.
In crowdsourcing, filtering out ``bad data'' is a constant battle with spammers, who can submit low-effort or nonsensical data and still receive compensation for their work \cite{DBLP:journals/corr/abs-1801-02546}.
In our incentive mechanisms, contributors do not benefit from submitting bad data and can even pay a penalty; meanwhile, honest contributors are actively incentivized to correct others' mistakes.

\subsection{Machine Learning and Blockchain Background}
We mainly considered supervised learning problems where a dataset consists of labeled samples.
For example, in a recommender system, a movie or restaurant is given a label from $1$ to $5$ stars.
The term \emph{model} refers to a machine learning algorithm that has been trained on data.
It is used to make predictions, e.g. predict the label of a given example.
It can be represented as a neural network, a matrix of numbers, a decision tree, etc.


Our framework applies to platforms where decentralized networks agree on a shared sequence of computations.
An example is the Ethereum blockchain \cite{buterin2015ethereum}.
A \emph{smart contract} is an object (in the sense of object-oriented programming) in this shared code.
It contains data fields and interacts with new code and events via its method calls.
A computation \emph{on-chain} means the computation is done \emph{inside of} a smart contract.
The input and result of the computation are usually stored on the blockchain.
In contrast, \emph{off-chain} means the computation can be done locally on the client's machine and does not necessarily need to be public.

In Ethereum, reading and running code provided by a smart contract has no cost if it does not write to the blockchain.
This means that one can use the model in a smart contract for inference for free.
When we discuss blockchains, smart contracts, and examples throughout this paper, we are mainly referring to Ethereum blockchain and the specifics of smart contracts on the Ethereum platform.
However, this design is certainly not limited to only run on Ethereum.

\subsection{Staking a Deposit}
\label{section:staking}
In conventional legal systems, violating an agreement may result in a penalty or fine.
Enforcing a penalty via a smart contract is complicated because a user cannot be forced to make a payment.
Instead, many solutions in the blockchain space require users to ``stake'' deposits that can be re-claimed later if they obey rules.\footnote{E.g. many solutions using Proof-of-Stake (PoS) such as in Tendermint \cite{kwon2014tendermint} and the Casper system for Ethereum \cite{buterin2017casper} involve staking a deposit to be considered eligible to participate.}
Similarly to those systems, we will also propose staking a deposit to simplify some incentive mechanisms for submitting new data.

\subsection{Organization}
In section \ref{sec:ml_models}, we describe the machine learning data handler and model portion of the algorithm.
In section \ref{sec:incentives}, we give our incentive mechanisms.
Then we discuss some implementation details in section \ref{sec:implementation}.
In section \ref{sec:benefits_of_blockchain}, we present some reasons for using a blockchain for this framework.
We present our responses to several potential issues in section \ref{sec:issues}.
Finally, we motivate some future work in section \ref{sec:future_work}.

\section{Machine Learning Models} \label{sec:ml_models}

This system can be used with various types of models including supervised and unsupervised.
The model architecture (e.g. different types of neural networks, CRFs \cite{lafferty2001conditional}, SVMs \cite{Cortes1995SVM}, etc.) chosen relates closely to the incentive mechanism chosen.
In our examples, we mainly consider training supervised classifiers because they can be used for many applications.
We first propose to leverage the work in the Incremental Learning space \cite{Schlimmer:1986:CSI:2887770.2887853} by using models capable of efficiently updating with one sample.
This should lower the transaction costs (``gas'') to update a model hosted in an Ethereum smart contract because each data contribution will be small.
One simple model with ``cheap'' updates is a Nearest Centroid Classifier, also known as a Rocchio Classifier \cite{manning2008rocchio}.
Online learning algorithms could also be appropriate.

\subsection{Initial Model}
It is often helpful if the initially deployed model has been trained to some extent.
For example, if the model is already somewhat useful, it is more likely to receive users for inference, who are more likely to generate and provide back data to improve the model.
Also, for some but not all of our incentive mechanisms, having an initial level of accuracy allows better validation of contributed data.
We stress that although this can be helpful, in many cases the model does not need to be pre-trained or highly accurate.

\subsection{Limitations}
\label{sec:limitations}
Due to limitations such as the cost of memory in the Ethereum blockchain, our examples usually focus on applications related to handling small input that can be compressed easily, such as text.
Text can easily be compressed using vocabulary dictionaries or with common shared encoders such as the Universal Sentence Encoder in \cite{cer2018universal}.
Complex models such as deep neural networks capable of processing images may be costly to store and update using Ethereum.
Just uploading the raw images would be costly.
For example, uploading all of the popular MNIST dataset of handwritten digits would cost about 275 ether at a modest gas price of 4gwei according to \cite{danku_protocol}.
At the time of this writing (May 2019) that is about 65,000USD.
Of course this cost would be amortized amongst all data contributors.

The transactions could incur high gas costs especially for the initial model deployment if the model or smart contract is large.
With high gas costs it is possible that nodes will reject the transaction.
Currently (May 2019) Ethereum has a limit of around 8 million gas.
In our experiments 8M gas is enough to deploy simple models and to submit complex data samples.
Even if the entire model is too large to submit when the contract is first deployed, the model can be added to the contract in several transactions after the contract is created.

\subsection{Experiment}
\begin{table}[b!]
    \caption{Gas Costs}
    \label{table:deployment_gas_costs}
    \begin{center}
        \begin{tabular}{|c|c|c|}
            \hline
            \textbf{Action} & \textbf{Gas Cost} & \textbf{USD$^{\mathrm{b}}$} \\
            \hline
            Deploy model contract & 3,845,840 & \$4.06 \\
            \hline
            Add data with 15 words ($h(x) = y$)$^{\mathrm{a}}$ & 177,693 & \$0.19 \\
            \hline
            Add data with 15 words ($h(x) \neq y$)$^{\mathrm{a}}$ & 249,037 & \$0.26 \\
            \hline
            \multicolumn{3}{l}{$^{\mathrm{a}}$Perceptron models are only updated when $h(x) \neq y$.} \\ \multicolumn{3}{l}{$^{\mathrm{b}}$In May 2019 with a modest gas price of 4gwei.}
        \end{tabular}
    \end{center}
\end{table}

We trained a single layer perceptron model \cite{rosenblatt1958perceptron} on the IMDB reviews dataset introduced in \cite{maas-EtAl:2011:ACL-HLT2011}.
The model has 100 binary features which are the presence of the 100 common words used in the dataset.
Thus 100 weights are initially deployed with the model.
Table \ref{table:deployment_gas_costs} shows the gas cost for various interactions with the model, $h$, using the Ethereum blockchain.
When data is added, $x$ is the features and $y$ is the label assigned to $x$.
The gas costs for adding data are calculated by adding data via the main contract entry point (not directly to the model contract).
When using more features, initial deployment cost increases but the cost to add data is similar because the features in data have a sparse representation.

\section{Incentive Mechanisms}
\label{sec:incentives}
The proposed incentive mechanisms (IM) encourage contributors to submit data that will improve the model's accuracy.
This can be measured in various ways.
The most natural proxy is to measure the performance with respect to a specific test set.
We will also discuss ways to measure performance if a test set cannot be provided.
The purpose of this paper is to present the general framework with motivating examples, as such, each incentive mechanism will be analysed further in future work.

We refer to \emph{good data} in the following sections as data that is objectively correct.
E.g. for a picture of the number 1, the label ``1'' is clearly better than the label ``0''.
Data can also be \emph{bad} (i.e. wrong) or \emph{ambiguous}.

\subsection{Gamification}
We first propose a baseline with no financial incentives in any form with the goal of reducing the barrier to entry.
This is the Wikipedia \cite{wikipedia} for Models \& Datasets.
This proposal relies solely on the willingness of contributors to collaborate for free, for a common good.

Additionally, points and optionally badges can be awarded to data contributors, i.e. stackexchangification \cite{stackexchange}. Badges in Stack Exchange have been shown to be effective by \cite{FM7299}.
The points and badges can be recorded on-chain in a smart contract using the contributor's wallet address as a key or identification.
Here are some examples of measurements for awarding points or badges to a user:
\begin{itemize}
    \item a specified number of data samples has been contributed
    \item submitting diverse data samples
    \item submitting data with different labels
    \item submitting data routinely (e.g. weekly)
\end{itemize}

Further experiments must be done into how these metrics can be explicitly computed efficiently on-chain or expanded off-chain.

\subsection{Rewards Mechanism Based on Prediction Markets}
\label{bounty}

In this section, we describe a monetary reward-based system for incentivizing contribution of correct data.
This design extends proposals of \cite{abernethy2011collaborative} and \cite{waggoner2015market} for collaborative machine learning contests.
An outside party, such as an academic institution or a company, provides (1) a pool of reward funds and (2) a test dataset.
Participants are rewarded according to how well they improve the model's performance as measured by the test data.

When this provider is available, we will be able to give very robust incentives for participation.
The mechanism is also resilient against manipulative or malicious providers and participants.
For cases where there is no outside party, we suggest the mechanism in section \ref{section:DRT}.

\subsubsection{Overview}
The mechanism is given in Fig. \ref{fig:bounty-mechanism}.
There are three phases. In the commitment phase, the provider deposits the bounty and defines a loss function $L(h,D)$.
This is a measure of loss (or a surrogate or proxy metric) of any model $h$ on any dataset $D$ (typically the average loss on points in the dataset).
Finally, the provider cryptographically commits to a test dataset, a small random fraction of which is initially revealed, similar to how data is revealed in \cite{danku_protocol}.

In the participation phase, people add data or otherwise provide updates to the model.
Each participant is required to deposit or ``stake'' $1$ unit of currency along with their update.\footnote{For the purposes of this description, each interaction is considered a separate participant. This is useful because participants cannot gain anything by creating false identities.}
After an end condition is met (such as a maximum time limit or amount of data), this phase ends.
A new cycle can begin if a new provider decides to commit new test data.

In the reward phase, the provider uploads the test dataset and the smart contract checks that it satisfies the commitment.\footnote{We note that, because the test data is not uploaded until the end, participants do not see the current performance of the model on the final test set and cannot overfit to it.}
Then, the smart contract determines rewards for all participants, as discussed next.

\begin{figure}
\begin{algorithmic}[1]
  \Statex \textbf{A. Commitment Phase}
    \State Provider deposits $B$ units of currency.
    \State Provider defines a loss function $L(h,D)$.
    \State Provider secretly divides a test dataset into $100$ equal parts and uploads their $100$ cryptographic hashes.
    \State Smart contract randomly selects $10$ of these hashes.
    \State Provider uploads $10$ partial datasets. If they do not match the $10$ hashes, abort.
    \State Provider specifies end condition (e.g. time limit).
\end{algorithmic}
\begin{algorithmic}[1]
  \Statex \textbf{B. Participation Phase}
    \State Smart contract contains an initial model, $h_0$.
    \For{each participant $t=1,2,\dots,T$ until end condition is met:}
      \State Participant deposits a stake of $1$ unit of currency
      \State Participant provides data.
      \State Model is updated from $h_{t-1}$ to $h_t$.
    \EndFor
\end{algorithmic}
\begin{algorithmic}[1]
  \Statex \textbf{C. Reward Phase}
    \State Provider uploads $90$ partial datasets; call them $D$. If they do not match the remaining $90$ hashes, abort.
    \State Let $b_t = 1$ for all $t$  \Comment{ balance initially equals stake}
    \State Let list $S = (1,\dots,T)$  \Comment{ list initially contains everyone}
    \For{$i=1,\dots,B$}
      \For{each participant $t$ in $S$}
        \State Let $t'$ be previous participant in $S$, or $0$ if none.
        \State Participant $t$'s balance is changed:
          \[ b_t \leftarrow b_t + L(h_{t'}, D) - L(h_t, D) \]
      \EndFor
      \State Let list $S = (t \in S : b_t \geq 1)$.  \Comment{ all who can re-stake $1$ stay in $S$}
    \EndFor
    \State Each participant $t$ is paid $b_t$.
\end{algorithmic}
\caption{\textbf{Bounty-based Incentive Mechanism.} We use $h$ to denote a machine learning model and $D$ for a dataset. The loss function $L(h,D)$ is clipped to the range $[0,1]$ by the smart contract.}
\label{fig:bounty-mechanism}
\end{figure}

\subsubsection{Reward calculation.}
First imagine that the bounty $B=1$, so that each participant $t$'s reward is their stake plus the following number:
\begin{equation}
  L(h_{t-1}, D) - L(h_t, D) .  \label{eq:loss-diff}
\end{equation}
This is exactly the reward function proposed in \cite{abernethy2011collaborative}, based on automated-market-maker or scoring-rule based prediction markets~\cite{hanson2003combinatorial}.
It can be pictured as follows: The smart contract first pays $L(h_0,D)$ to the first participant.
Their data updated the model to $h_1$, so they pay $L(h_1,D)$ to the second participant.
This continues down the line until the last participant pays $L(h_T,D)$ back to the smart contract.
The better $h_t$ performs, the less participant $t$ has to pay forward, so they are incentivized to provide data that is as useful as possible relative to the (expected) test set.
(If $h_t$ performs worse than the previous model, $t$ loses some or all of their stake.)
In total, the smart contract pays out a net amount of $L(h_0,D) - L(h_T,D)$, which is the total improvement from all contributions.
It is at most $1$ by assumption on the loss function.

Finally, we must scale this mechanism for a bounty of $B \gg 1$.
The approach of \cite{abernethy2011collaborative} would require that all participants stake $B$, which is infeasible.
Therefore, instead, we use the approach of iterating the mechanism $B$ times.
Each iteration, the participant stakes $1$ unit, then receives a reward.
If she can no longer stake $1$ unit due to losses, she drops out.\footnote{Of course, our mechanism can be modified to allow participants to stake more than $1$ unit in order to cover more losses, but it is not clear if this would be beneficial to them.}
Although this is slightly complex, it still remains that the better $h_t$, the more reward $t$ gets, so we believe incentives for participation are strongly aligned.

\subsubsection{Untrusted provider, commitment, and value burning}
The provider can potentially manipulate by first, the choice of dataset, and second, by participating or partnering with some participants.
As we describe next, the defenses against manipulation are the cryptographic commitment scheme along with value ``burning'', which occurs when some value that was deposited to the smart contract is not returned to anyone.

The cryptographic commitment scheme forces the provider to reveal in advance a random $10\%$ of the dataset (of course, the numbers of $10$ and $100$ can be adjusted as desired).
If the provider does not comply, the process is aborted and the reward $B$ stays stuck in the contract (not refunded).
Assuming the provider complies, participants learn something about the final dataset.
This prevents the following problematic attack: the provider secretly chooses a dataset with incorrect labels, then participates anonymously with corresponding updates.
The provider could then not only gain back most of the original reward, but also earn significant amounts from the stakes of the honest participants, which they would lose because their data is not helpful for this test set.
The commitment scheme reveals a representative part of the test set up front, so that participants would likely be alerted to such an attack and refuse to participate.

At the end of the process, a significant amount of the original bounty is likely to remain stuck in the contract.
Specifically, only the following amount is distributed:
\begin{equation}
  B \cdot \left[ L(h_0, D) - L(h_T, D) \right]  . \label{eq:total-bounty-paid}
\end{equation}
While this is not ideal, the benefit is that there is no incentive for the provider to choose an incorrect dataset in hopes of having significant value left over.
The smart contract could instead donate left over funds to some predetermined account (such as a charity), but that would be open to manipulation.

In summary, the provider must expect to lose the entire bounty.
Therefore, it will only participate if the benefit from the trained model is worth this cost, so that only honest providers have an incentive to join.\footnote{A colleague points out that a provider may be able to gain back some of the bounty by participating using the portion of the initially revealed test set. However, this is not harmful to the final performance of the model.}

\subsection{Deposit, Refund, and Take: Self-Assessment}
\label{section:DRT}
Ideally, one could enforce a fine or penalty on those submitting bad data.
One way to determine if data is bad is to have other contributors validate it as is common in conventional crowdsourcing methods.
However, enforcing a penalty at a later time via a smart contract is difficult as established in section \ref{section:staking}.
To facilitate penalties, this proposal enforces a deposit when contributing data.

This IM is an alternative where one does not need to rely on a benevolent agent to submit a test set as described previously.
Instead it is possible to rely on the data contributors to indirectly validate and pay each other.
As a proxy to verify data, we propose using the deployed model, $h$, to validate new contributions.
The caveat is that the initially deployed model needs to already be trained and generally already correctly classify samples with some accepted degree of accuracy.

Here are the highlights of the proposal:
\begin{itemize}
    \item \emph{Deploy}
    a model, $h$, already trained with some data.
    \item \emph{Deposit}:
    Each data contribution with data $x$ and label $y$ also requires a deposit, $d$.
    Data and meta-data for each contribution is stored in the data handler.
    \item \emph{Refund}:
    To claim a refund on their deposit, after a time $t$ has passed and if the current model, $h$, still agrees with the originally submitted classification, i.e. if $h(x) == y$, then the contributor can have their entire deposit $d$ returned.
    \begin{itemize}
        \item We now assume that $(x, y)$ is ``good'' data.
        \item The successful return of the deposit should be recorded in a tally of points for the wallet address.
    \end{itemize}
    \item \emph{Take}:
    A contributor that has already had data validated in the \emph{Refund} stage can locate a data point $(x,y)$ for which $h(x) \neq y$ and request to take a portion of the deposit, $d$, originally given when $(x,y)$ was submitted.
\end{itemize}

If the sample submitted, $(x,y)$ is incorrect or invalid, then within time $t$, other contributors should submit $(x, y')$ where $y'$ is the correct or at least generally preferred label for $x$ and $y' \neq y$.
This is similar to how one generally expects bad edits to popular Wikipedia articles to be corrected in a timely manner.


\subsubsection{Time to Wait for Refund}
The creator of the contract must select, $t$, how much time contributors need to wait before they can claim a refund on their deposit.
As a guideline, we propose setting $t$ to be enough time for other contributors to submit corrections with different labels if they disagree with the data.
For example, $t \geq \text{one week}$.

Models that are not very sensitive might need to allow more time to pass in order for enough samples to be provided to teach the model about a new use case.

Very sensitive models could allow malicious contributors to claim refunds for ``bad'' data before another contributor has a chance to ``correct'' their bad submission.
Such models should also require a deposit high enough to dissuade bursts of malicious data submissions.
Special care needs to be taken and experiments should be done before setting $t$.

Certainly $t$ does not have to be constant.
It could somehow depend on the provided data sample, frequency of data submitted, or even the certainty of the model on the data point.
I.e. if the model has a measure of probability of correctness, $P(h(x)=y)$, for submission $(x,y)$ then it can be used to reduce $t$ because it's unlikely to be changed later.
\begin{equation*}
    t \propto \frac{1}{P(h(x)=y)}
\end{equation*}

\subsubsection{Varying Deposit}
Requiring a deposit has a few goals:
\begin{itemize}
    \item Introduce value to the system allowing others the chance to profit after they have contributed correct data.
    \item Inhibit too frequent changes to the model.
    \item Reduce spam (incorrect or invalid data).
\end{itemize}

To achieve these goals we propose
\begin{equation*}
    d \propto \frac{1}{\text{time since last update}}
\end{equation*}
I.e. it is costly for contributors to send many updates in a short amount of time.
This should give those using the prediction function of the models a more consistent experience.
E.g. consider a personal assistant in one's home that behaves too differently to the same spoken request uttered several times a day, such as a request to play the news.

\subsubsection{Taking Another's Deposit}
We introduce some guidelines for a contributor that is reporting ``bad'' data to take some of the deposit from the original contributor, $c$.
Note that contributed data and meta-data about it can be found in the data handler or emitted events.

First some definitions:
\begin{itemize}
    \item
    Let $r(c_r, d)$ be the reward that the contributing reporter, $c_r$ receives for reporting data $(x,y)$ with deposit $d$.
    \item
    Let $n(c)$ be the number of data samples for which contributor $c$ received a \emph{refund} (assumed good data).
\end{itemize}

Guidelines:
\begin{itemize}
    \item $h(x) \neq y$: The current model disagrees with the label. So we \emph{assume} the data is ``bad''.
    \item $n(c_r) > 0$: The reporter should have already had data refunded.
    This enforces that they hopefully already submitted ``good'' data before they can try to profit from the system.
    \item $c_r \neq c$: The reporter cannot be the original contributor.
    Otherwise contributors can easily attempt to reclaim their deposit for ``bad'' data.
    \item The reward should be \emph{shared} amongst ``good'' contributors.
    \begin{equation} \label{eq:drt_reward_split}
        r(c_r, d) \propto d \times \frac{n(c_r)}{\sum_{\text{all } c} n(c)}
    \end{equation}
    \begin{itemize}
        \item This protects against Sybil attacks where a contributor can use a second account to take back their entire deposit.
        They can still claim back some of their reward from another account but  they will have to wait and get refunded for some ``good'' data using that other account.
    \end{itemize}
    \item $r(c_r, d) > \epsilon > 0$: The reward should be at least some minimal value to cover potential transaction costs.
    \item The data handler must keep track of the remaining deposit that can be claimed, $d_r \leq d$.
    $$ d_r \leftarrow d_r - r(c_r, d)$$
    $$ r(c_r, d) \leq d_r \leq d$$   
    \item Since $n(c)$ changes over time, the ratio in \eqref{eq:drt_reward_split} changes while reporters are claiming their share of $d$.
    Therefore, it possible that some reporters get a smaller proportion of $d$.
    We discuss some possible solutions to this in \ref{section:preventing_lock-ups}.
\end{itemize}

\subsubsection{Biasing the Model}
With the proposal, contributors can be tempted to only submit data the model already agrees with ($h(x)=y$) at submission time and hope the model still agrees with at refund time.
This could create a bias in the model towards data it already ``knows''.
Contributors would normally still have to pay a transaction fee so in effect they still pay a nominal fee to deposit and claim their refund.
The model and training method must be carefully chosen and the designer can consider how to handle duplicate or similar data.
The IM can even reject too much duplicate or similar data.

\subsubsection{Preventing Lock-ups}
\label{section:preventing_lock-ups}
In this section we discuss ways to avoid funds getting ``locked-up`` or ``stuck inside'' the smart contract.
It is possible that contributors omit to collect their refunds or that contributors do not take their portion of the deposit leaving value ``stuck inside'' the contract.
To avoid this we introduce two new parameters:
\begin{itemize}
    \item $t_c$: The amount of time the creator has to wait to take the entire remaining refund ($d_r$) for a specific contribution.
    Where $t_c > t$.
    Additionally, this gives creators some incentive to deploy a model as they \emph{may} get a chance to claim a significant portion of $d$.
    Contracts may want to enforce that this is much greater than the amount of time to wait for attempting a refund, which gives contributors even more time to get the deposit back and not allow the creator take too much ($t_c \gg t$).
    \item $t_a$: The amount of time anyone has to wait to take the entire remaining refund ($d_r$).
    Where $t_a \geq t_c > t$.
    in case the creator omits to take ``stuck'' value from the contract.
\end{itemize}

Certainly there can be more variants of these such as a value, $t_d$, for data contributors with refunded submissions ($n(c) > 0$) where $t_a \geq t_d \geq t_c$.

\subsubsection{Experiment}
\begin{figure}[b]
    \centerline{\includegraphics[width=0.48\textwidth]{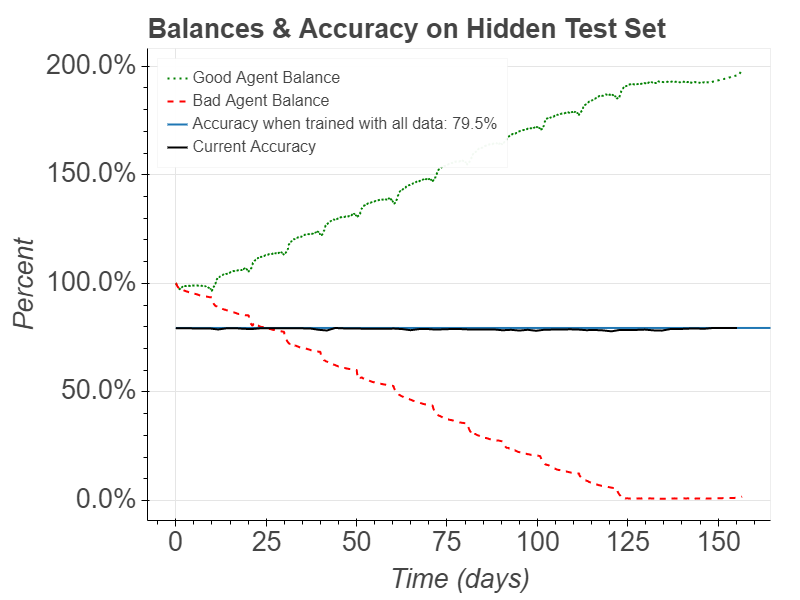}}
    \caption{Balance percentages and model accuracy in a simulation where an adversary, "Bad Agent" is willing to spend about twice as much on deposits than an honest contributor, "Good Agent". The adversary is only submitting data about one sixth as often. Despite the malicious efforts, the accuracy can still be maintained and the honest contributors profit. }
    \label{fig:drt_bal_acc}
\end{figure}

We developed simulations to test parameters for incentive mechanisms and models.
For one simulation, we initially trained a Perceptron model \cite{rosenblatt1958perceptron} with 8\% of the IMDB reviews training dataset introduced in \cite{maas-EtAl:2011:ACL-HLT2011}.
The model has 1000 binary features which are the presence of the 1000 most frequent words in the dataset.
Fig \ref{fig:drt_bal_acc} shows the results of a simulation where for simplicity, we show just one honest contributor and one malicious contributor but these contributors effectively represent many contributors submitting the remaining 92\% of the training data over time.
Details for the simulation can be found with our source code.

\section{Implementation}
\label{sec:implementation}
In this section we discuss some implementation details for our proposed framework.

\subsection{Floating Point Numbers}
In our examples we use integers for data representations because Ethereum does not support floating point numbers.
Whenever we expect a floating point number, we take in an integer that has already been multiplied by some value, e.g. $10^9$ (9 decimal places of precision).
All operations are done considering this transformation.

\subsection{Inversion of Control}
\begin{figure}[b]
    \begin{mdframed}
        \begin{center}
            \includegraphics[width=\textwidth,trim=18 18 18 18,clip]
                {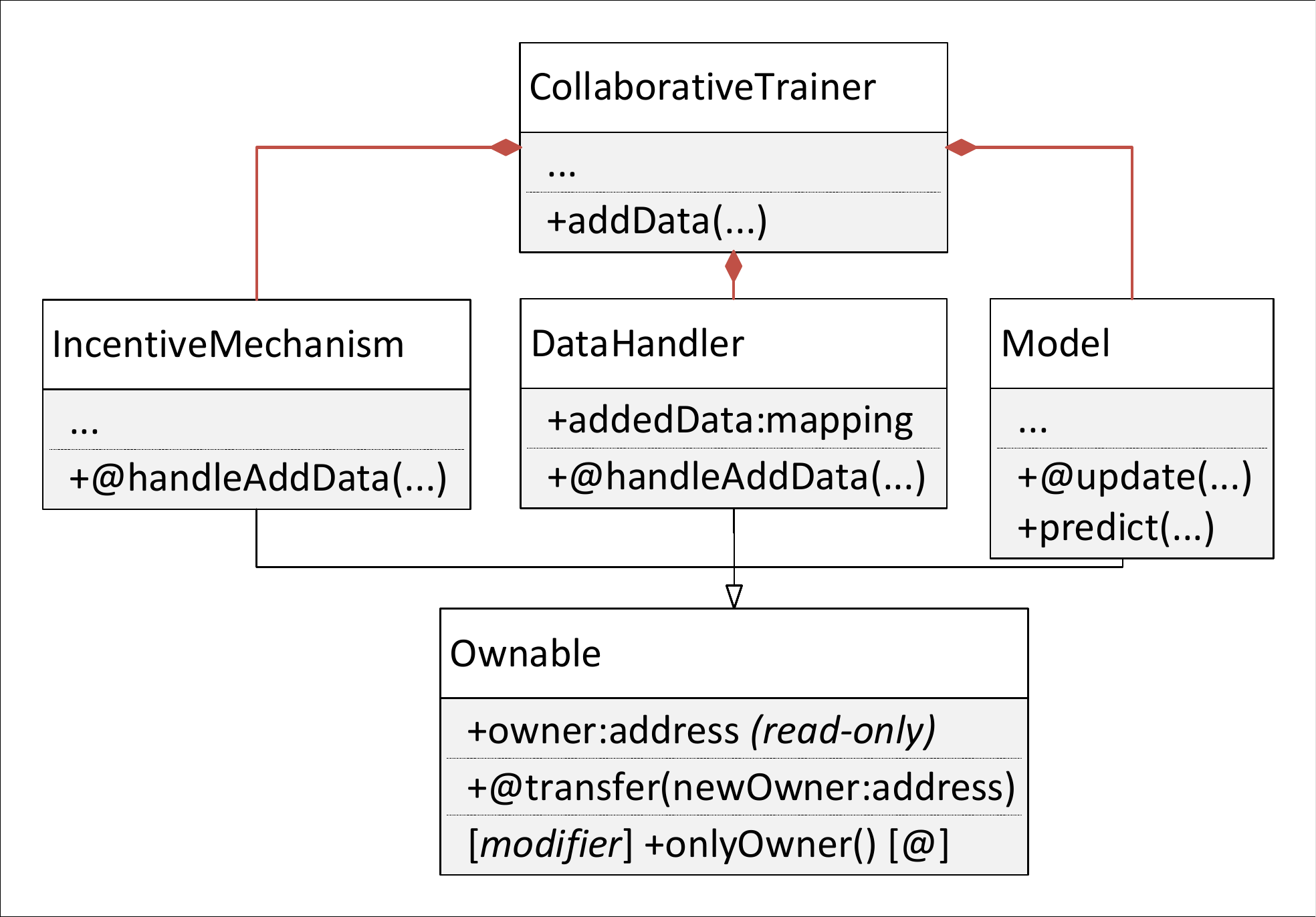}
        \end{center}
    \end{mdframed}
    \caption{Overview of the class diagram for the framework; other members and methods exist. We use ``@'' before a method to indicate that only the owner may call the method.}
    \label{fig:architecture}
\end{figure}

To favor clarity and ease development, our examples use the Inversion of Control \cite{wikipedia:inversion_of_control} principle favoring composition over inheritance.
In Ethereum smart contracts using inheritance can be cheaper than ones using composition according to \cite{ethereumstackexchange:composition_over_inheritance}.
Some subtleties such as ownership and publicity of contained contracts need to be considered.
Fig. \ref{fig:architecture} shows a class diagram for the proposed framework.
\iftoggle{detailed_version}{%
}{%
}

Since the model will exist as its own contract it will need to have its update method exposed publicly.
Only the owner of the model (the CollaborativeTrainer) can call the model's update method because this owner is responsible for using the incentive mechanism.
\iftoggle{detailed_version}{%
    The owner defaults to the creator of the smart contract.
    Once the main contract is created the creator must transfer ownership of the model and incentive mechanism to the main contract so that the model can only be trained by interfacing with the main contract.
}{%
}

\section{Benefits of Blockchain}
\label{sec:benefits_of_blockchain}

Using a blockchain (such as Ethereum \cite{buterin2015ethereum}) provides us with many advantages compared to traditional source code hosting and deployment:

\subsection{Public, Persistent \& Decentralized}
The model is a shared and publicly visible resource.
One can essentially trust that the model will persist and be available on one of many decentralized nodes.

\subsection{Versioning}
It is easy to revert to an earlier version of the model by only updating your node's version of the blockchain up to the particular block where your application is satisfied with the model, possibly measuring performance with respect to one's own hidden test set.
\iftoggle{detailed_version}{%
    With constructs like ``events'' on the Ethereum blockchain, it is possible to get a specific version of the model and the data contributed.
}{%
}

\subsection{Models Evolve Over Time}
Smart contracts on a blockchain allow models to evolve as the environment changes, for example recognize novel words.

\subsection{Transparency \& Trust}
Typically a machine learning or data collection service would be set up with a REST API or some other system where users interface with code running on some servers.
Users can submit or request data from the servers but there is no way for the user can be completely certain of the source code running on the servers.
Code running in an Ethereum smart contract is completely public and the result of a particular execution can be deterministically evaluated.
Essentially, data contributors can trust that the model updates and that they will be compensated for their contributions.

\iftoggle{detailed_version}{%
    \subsection{Payment}
    One does not need a bank account to participate and as such payments are simplified compared to typical crowdsourcing methods.
    By avoiding payment through typical financial institutions it is possible contributors have less fees to pay when making deposits or receiving compensation compared to a system that uses fiat to pay contributors.
    This depends on several factors such as one's bank and if one needs to eventually convert the blockchain's currency into fiat.
}{%
}

\section{Potential Issues}
\label{sec:issues}
Data contributors and smart contract creators should consider several vulnerabilities when using our framework.
Many possible issues have already been addressed for systems with static models in \cite{danku_protocol}.
This section analyzes issues specific to systems where models can be trained.

\subsection{Submitting Bad Data}
\label{sec:submitting_bad_data}
A wealthy and determined agent can corrupt a model deployed by submitting incorrect or nonsensical training data.
\paragraph*{Response}
The incentive mechanism should make it costly and unbeneficial to submit bad data.
Furthermore, many blockchains have transactions fees making it very costly to submit a lot of data.
Even if a model is corrupted, the users of the model can just revert back to an earlier version of the model since it is on a blockchain.
Additionally, analysis can be done to salvage ``good'' data already submitted.

\subsection{Ambiguous Data}
\label{ambiguous_data}
Those using this framework must carefully consider the type of model, IM, and how submitting ambiguous data can affect the system.
For example, the sentiment of ``The movie is okay'' if the options for the label are ``happy'' or ``sad''.
Ambiguous data is always an issue when crowdsourcing but is especially concerning here since contributors can lose funds.
It is safest for contributors to keep their data unambiguous.



\subsection{Overwhelming the Network}
Some applications relying on public blockchains have had reliability issues due to network congestion.
This can be an issue for this framework when adding data which requires creating new transactions.
Inference should not be affected because running inference just involves reading which is normally free and can be done locally with the latest local version of the model.

\subsection{Requiring a Deposit}
Many new contributors might not be familiar with the established practice of blockchain applications requiring a deposit.
Small fees are already required to process any transaction in most of the popular blockchains such as Bitcoin \cite{nakamoto2008bitcoin} and Ethereum \cite{buterin2015ethereum}.

There are some ways to hide these fees from end users.
A third-party can hide the deposit cost by providing their own interface to a public system and validating data contributions themselves.
Perhaps this third party believes they have algorithms and better means to validate data.
They could then reward users under their own scheme which may not even involve financial incentives.

\section{Future Work}
\label{sec:future_work}
There are a few areas where the presented framework can be configured and built upon.

\subsection{Models}
More research needs to be done on the types of models that will work well within this framework.

\subsubsection{Unsupervised Models}
The examples discussed mainly use supervised classifiers because we focus on validating data with labels.
Test sets and incentive mechanisms to validate the data contributions can still be used with unsupervised models.
Some examples are:
\begin{itemize}
    \item Clustering: Developing clustering models especially for outlier detection can be very useful.
    \item Generative models such as autoencoders and GANs.
        For example, a model could be built that attempts to generate text or draw pictures.
\end{itemize}

\subsubsection{Complex Models}
\label{sec:complex_models}
There are cost limitations as described in section \ref{sec:limitations}.
We propose more research in pre-computing as much as possible off-chain and only performing necessary steps in a smart contract.
Techniques such as \emph{fine-tuning} (in the transfer learning field) \cite{bengio2012deep} or off-chain training as proposed for DeepChain \cite{Weng2018DeepChainAA} can help.
One can use a common encoder (shared in more conventional ways such as via web APIs or publicly posting source code to a website) \emph{off-chain} to encode the input and then fine-tune the encoded result \emph{on-chain} with a simple neural network or even just a single layer.
While the encoder should be static, it is possible for it to change slightly as long as the input to the fine-tuning component is similar enough to previously seen training data.


Complex models can even be interfaced via an API through the smart contract using a system such as Provable (formerly Oraclize) \cite{provable}.
Hiding the model behind an API means the model is not necessarily public which is not in the spirit of this proposal.
Complex models can also be built up by combining several ``layers'' of smart contracts using this framework.

\subsubsection{Recovering Corrupted Models}
Work can be done in how to recover a model corrupted by bad data (as described in section \ref{sec:submitting_bad_data}).
Once a dataset is collected it can be further refined through various methods (e.g. clustering data).
The cleaned dataset can be used to train a new model.
This new model could be kept private by those that organize cleaning efforts and used in their production system.
The model could also be used to start a new instance of the collaborative training process and collect more training data.


\subsection{Incentive Mechanisms}
More exploration, analysis, and experiments with incentive mechanisms in this space needs to be done with emphasis on the type of model each incentive mechanism works well with.

The incentive mechanisms imposed by the smart contract could be hidden to end users by 3rd party services that build services around this proposed framework.
These services could validate data contribution themselves offering their own rewards to users of their platforms that do not wish to interact with these smart contracts.

\subsection{Privacy}
Contributors may not want to publish their data to a public blockchain.
Initially we propose to only use this for framework for data that is safe to become public.
E.g. certain queries to a personal assistant such as, ``What will the weather be like tomorrow?'', which contains no personal data.

\iftoggle{detailed_version}{
    For private data, Homomorphic Encryption techniques such as the tools in \cite{sealcrypto} may help if the model works within those constraints.
}
Future work can be done to not submit data directly to the smart contract and instead just submit model updates as discussed in section \ref{sec:complex_models} and in DeepChain \cite{Weng2018DeepChainAA}.

\section{Conclusion}
\label{section:conclusion}
We have presented a configurable framework for training a model and collecting data on a blockchain by leveraging several baseline incentive mechanisms and existing types of machine learning models for incremental learning.
Ideal scenarios have varying data with generally agreed upon labels.
Currently, this framework is mainly designed for models that can be efficiently updated but we hope to see future research in scaling to more complex models.

\section*{Acknowledgement}
The authors would like to thank William Buchwalter for believing in the ideas initially and helping to implement the first demos.
Nicole Immorlica, Brendan Lucier, Dave Pennock, and Glen Weyl for discussions and suggestions.
Adam Atkinson, Hannes Schulz, Kaheer Suleman, and Jaclyn Hearnden for giving detailed feedback on the paper.


\end{document}